\documentclass[a4paper,11pt]{article}
\pdfoutput=1 

\usepackage{jheppub} 
\usepackage{physics}

\usepackage[T1]{fontenc} 

\preprint{OU-HET-1053}

\title{Islands in Schwarzschild black holes}

\author{Koji Hashimoto,}
\author{Norihiro Iizuka, and} 
\author{Yoshinori Matsuo}

\affiliation{Department of Physics, Osaka University,\\1-1 Machikaneyama, Toyonaka, Osaka 560-0043, Japan}

\emailAdd{koji@phys.sci.osaka-u.ac.jp}
\emailAdd{iizuka@phys.sci.osaka-u.ac.jp}
\emailAdd{matsuo@het.phys.sci.osaka-u.ac.jp}

\abstract{
We study the Page curve for asymptotically flat eternal Schwarzschild black holes in four and higher spacetime dimensions. 
Before the Page time, the entanglement entropy grows linearly in time. After the Page time, the entanglement entropy of a given region outside the black hole is largely modified by the emergence of an island, which extends to the outer vicinity of the event horizon. As a result, 
it remains a constant value which reproduces the Bekenstein-Hawking entropy, consistent with the finiteness of the von Neumann entropy for an eternal black hole.   
}

\begin{document}

\maketitle
\flushbottom


\section{Introduction and our strategy}
\label{sec:intro}

The information paradox \cite{Hawking:1976ra} is the most fundamental problem in quantum gravity. 
The Hawking radiation behaves as thermal radiation \cite{Hawking:1974sw}, 
which implies that the entanglement entropy outside the black hole is monotonically increasing. 
On the other hand, 
quantum mechanics requires that 
the entanglement entropy goes to zero 
at the end of the evaporation
since it must be the pure state. 
The time evolution of the entanglement entropy is described by the so-called Page curve \cite{Page:1993wv,Page:2013dx}. 
Thus, the information loss paradox is translated to 
the problem how the Page curve is reproduced for 
the entanglement entropy of the Hawking radiation. 

Recently it was proposed that the Page curve emerges from the effect of {\em islands} 
\cite{Penington:2019npb,Almheiri:2019psf,Almheiri:2019hni,Almheiri:2019yqk}. 
Regarding the state of the Hawking radiation as
that in a region $R$ outside the black hole, 
the density matrix of $R$ is normally defined by taking the partial trace over 
the states in $\overline R$, which is the complementary region of $R$. 
According to the prescription of the minimal quantum extremal surface 
\cite{Ryu:2006bv,Hubeny:2007xt,Engelhardt:2014gca}, 
states in some regions in $\overline R$, which are called islands $I (\subset \overline R)$, 
should be excluded from the states to be traced out. 
Thus, the entanglement entropy of the Hawking radiation $R$ 
is effectively given by that of states in $R \cup I$. 
Explicitly, the entanglement entropy of the Hawking radiation is give by 
\begin{equation}
\label{GEformula}
 S(R) 
 = 
 \min \left\{\mathrm{ext}\left[
 \frac{\mathrm{Area}(\partial I)}{4G_{\rm N}} 
 + S_{\rm matter}(R\cup I)
 \right]\right\} \ , 
\end{equation}
by using the prescription of the quantum extremal surface.

The island rule was first proposed as a result of the conjectured quantum extremal surface prescription, 
and recently the island rule was derived by using the replica method for the gravitational path integral. 
When one applies the replica trick \cite{Callan:1994py,Holzhey:1994we,Calabrese:2009qy} 
to gravitational theories, 
one can fix only the boundary conditions of the replica geometries, 
and new saddles, where bulk wormholes are connecting different 
copies of spacetime, need to be taken into account. 
These new saddles, called  {\em replica wormholes}, lead to islands \cite{Penington:2019kki,Almheiri:2019qdq}. 
In the semi-classical limit of gravity, 
the partition function of the geometry with replicas 
is dominated by that giving the minimum entanglement entropy. 
In this way, the replica trick for gravitational theories  
leads to the same formula \eqref{GEformula} as the quantum extremal surface prescription.

Since the replica wormhole is merely a consequence of the replica trick 
in models with gravitation, the island conjecture is expected to be applicable to any black hole. 
So far, among recent works \cite{Penington:2019npb,Almheiri:2019psf,Almheiri:2019hni,Almheiri:2019yqk,
Penington:2019kki,Almheiri:2019qdq,
Chen:2019uhq,Chen:2019iro,Akers:2019lzs,Liu:2020gnp,Marolf:2020xie,Balasubramanian:2020hfs,
Bhattacharya:2020ymw,Verlinde:2020upt,Chen:2020wiq,Gautason:2020tmk,Anegawa:2020ezn}, 
the island rule has been studied mainly in two spacetime dimensions,\footnote{See ref.~\cite{Almheiri:2019psy} for the study of islands in higher dimensions.} which offer a
tractable treatment of the entanglement entropy of the Hawking radiation. 
In this paper, we make one more step for general black holes. 
We study the effect of islands in the Schwarzschild black holes, 
in asymptotically flat spacetime in generic dimensions. 
Needless to say, the asymptotically flat four-dimensional Schwarzschild black hole is the first solution historically \cite{Schwarzschild:1916uq} and the simplest and the most interesting black hole. We start with the four-dimensional case. 

The gravitational part of the action is given by the Einstein-Hilbert action with the Gibbons-Hawking term, 
\begin{align}
 I &= I_{\rm gravity} + I_{\rm matter} \, ,
 \\
 &I_{\rm gravity} 
 = 
 \frac{1}{16\pi G_{\rm N}} \int_{\mathcal M} d^4 x \sqrt{-g} \, R 
 + \frac{1}{8\pi G_{\rm N}} \int_{\partial \mathcal M} d^{3} x \sqrt{-h} \, K \ , 
\label{EinsteinHilbert}
\end{align}
where $G_{\rm N}$ is the Newton constant.\footnote{It is straightforward to generalize the analysis to gravity with higher curvature terms, but we focus only on the dominant contributions.}
Our goal in this paper is to show that the entanglement entropy of the Hawking radiation 
of the Schwarzschild black hole follows a Page curve once islands are taken into account. 
The Schwarzschild black hole metric we consider is 
\begin{align}
ds^2 = - \,  \frac{r - r_{\rm h}}{r} dt^2 + \frac{r}{r - r_{\rm h}} dr^2 + r^2 d \Omega^2 
\end{align}
with the horizon radius $r_{\rm h}$. Its temperature is 
\begin{align}
T_{\rm H} = \frac{1}{\beta} = \frac{1}{4 \pi r_{\rm h}}  \,.
\end{align}

In the following we summarize our analyses and necessary ingredients for them.
First, we will apply the quantum extremal surface (or equivalently, the replica wormhole) prescription to gravity theory with matter fields. 
We do not resort to holographic correspondences, nor to embedding into higher-dimensional (AdS) spacetime, nor 
to coupling with an auxiliary system to absorb the radiation. 
We will use the global two-sided geometry.

Before we proceed, we comment on the formula \eqref{GEformula}. 
The formula \eqref{GEformula} consists of two terms;  
the gravitational part%
\footnote{%
The gravitational part of the generalized entropy contains 
the term proportional to the area of $\partial R$. 
It comes from the effect that the region $R$ is separated from the other 
and is irrelevant to the entropy of the Hawking radiation. 
In fact, it exists even in the case of the empty flat spacetime (without black holes) as far as 
we consider bulk gravitational theories.  
} of the generalized entropy, $ S_{\rm gravity} $,
which is proportional to the total area (or volume for $D>4$) of the boundaries of an island, $\partial I$,  
 as \cite{Faulkner:2013ana,Lewkowycz:2013nqa,Dong:2016hjy,Dong:2017xht} 
\begin{equation}
\label{gravientropyarea}
 S_{\rm gravity} = \frac{{\rm Area}(\partial I)}{4 G_{\rm N}} \,,
\end{equation}
and the matter entanglement entropy $S_{\rm matter}$ on the region $I$ and $R$ in the curved background.  
Note that the formula given by eq.~\eqref{gravientropyarea} is consistent with our action \eqref{EinsteinHilbert}. 
Note also that without islands, the gravitational entropy of an island, $I$, vanishes. 

Unlike two-dimensions, in four-dimensions, it is well-known that the matter entanglement entropy has area-like divergences, which depend on the short distance cut-off \cite{Bombelli:1986rw, Srednicki:1993im}. 
Therefore, this yields the following divergence for the matter entropy  
\begin{equation}
\label{cut-off-divergence}
S_{\rm matter}(R\cup I) = \frac{{\rm Area}(\partial I)}{\epsilon^2} + S^{\rm (finite)}_{\rm matter}(R\cup I)  \,, 
\end{equation}
where $\epsilon$ is the short distance cut-off scale.\footnote{Similarly we have area-divergence coming from the boundary of $R$, which is irrelevant and we will neglect in this paper.} 
This divergence can be absorbed by the renormalization of the Newton constant as \cite{Susskind:1994sm}  
\begin{equation}
 \frac{1}{4 G^{(r)}_{\rm N}}=\frac{1}{4 G_{\rm N}} + \frac{1}{\epsilon^2} \,, 
\end{equation}
where ${G}_{\rm N}$ is bare Newton constant and $G^{(r)}_{\rm N}$ is {\it renormalized Newton constant}. 
In this respect, 
if we regard ${G}_{\rm N}$ in the formula \eqref{GEformula} as the renormalized Newton constant, then 
the leading cut-off dependent divergence of $S_{\rm matter}(R\cup I)$ in eq.~\eqref{cut-off-divergence}  is already taken into account, and therefore 
$S_{\rm matter}(R\cup I)$ yields only a {\it finite} contribution, {\it i.e.,} $S^{\rm (finite)}_{\rm matter}(R\cup I)$; Therefore our proposal formula in higher dimensions is 
\begin{equation}
\label{GEformula2}
 S(R) 
 = 
 \min \left\{\mathrm{ext}\left[
 \frac{\mathrm{Area}(\partial I)}{4G^{(r)}_{\rm N}} 
 + S^{\rm (finite)}_{\rm matter}(R\cup I)
 \right]\right\} \,. 
\end{equation}
By evaluating the formula \eqref{GEformula2} as the prescription of the minimal quantum extremal surface in higher dimensions, in this paper,  
we will derive the Page curve. 
For the finite matter entropy contribution, $S^{\rm (finite)}_{\rm matter}(R\cup I)$, we will use eq.~\eqref{SwoI} and eq.~\eqref{SwI}. 
This can be understood as follows.

The region for the Hawking radiation $R$ in the Schwarzschild spacetime is 
the union of two regions $R_+$ and $R_-$ 
which are located in 
the right and left wedges in the Penrose diagram, respectively (see Fig.~\ref{fig:BH}). 
The distance between $R_+$ and $R_-$ becomes very large at late times (see appendix~\ref{sec:app}),
therefore at late times, the entanglement entropy of the Hawking radiation 
without islands is expected to be very large and 
the configuration with islands is expected to give the dominant contribution. 

First, we consider the configuration without islands. The matter entanglement entropy we will evaluate is that of 
separated two regions $R_+$ and $R_-$ (see Fig.~\ref{fig:BH} Left).
In this case,  
the finite contributions of the matter entanglement entropy is, $S(R_+ \cup R_-)$ minus $S(R_+) + S(R_-)$,  
which is essentially the minus of the mutual information $I(R_+;R_-)$.\footnote{  
The mutual information $I(A;B)$ is given by, 
\begin{align}
 I(A;B) \equiv  - S(A \cup B)  + S(A) + S(B) \,. 
\end{align}}
This is because the leading contributions of the entanglement entropy of each region, $S(R_+)$ and $S(R_-)$, 
are the divergences of the form \eqref{cut-off-divergence}, proportional to the area of the boundary surface.    
Hence these cut-off dependent leading boundary-area divergences are already taken into account  
by the renormalized Newton constant and do not contribute to $S^{\rm (finite)}_{\rm matter}(R \cup I)$.%
\footnote{%
There are higher order corrections to the area terms of the entanglement entropy, 
which are also renormalized into the higher order gravitational constants. 
} 
Thus, in the case of no island, the finite part of the matter entanglement entropy is given 
in terms of the mutual entropy as 
\begin{align}
  S^{\rm (finite)}_{\rm matter}(R\cup I)  = S^{\rm (finite)}_{\rm matter}(R)  = - I (R_+;R_-)  \qquad  \mbox{(without island)}
 \label{SwoI}
\end{align}

Next, we consider the configuration with an island $I$. At late times, 
each of two boundaries of $I$ is much closer to 
the boundary 
of $R$ in the same wedge, 
than to the boundaries of $R$ and $I$ in the other wedge (see Fig.~\ref{fig:BH} Right).%
\footnote{%
The distance between the right and the left boundaries is characterized by the volume of an extremal surface
connecting the boundaries. The calculation of that extremal surface resembles that of a holographic complexity 
(``complexity = volume'' conjecture \cite{Stanford:2014jda,Susskind:2014jwa}). As shown in appendix~\ref{sec:app}, at late times, the volume of the three-dimensional extremal surface
in the analytically continued Schwarzschild geometry grows linearly in time. The section of this extremal surface
is at most $4 \pi b^2$ where $b$ is the value of the radial coordinate of the Schwarzschild geometry, thus 
the extremal three-dimensional surface is a long cylinder at late times. This means that the four-dimensional matter free fields can be 
treated as a two-dimensional massless fields which are the lowest mode in the KK towers.
}  
The correlation between the left and right wedges 
is negligible since the neighboring boundary (hyper)surfaces of different regions 
behaves like charges with opposite sign. 
Thus the total entanglement entropy is well dominated by 
that in each side separately. 
\begin{figure}[t]
\centering
	\includegraphics[width=70mm]{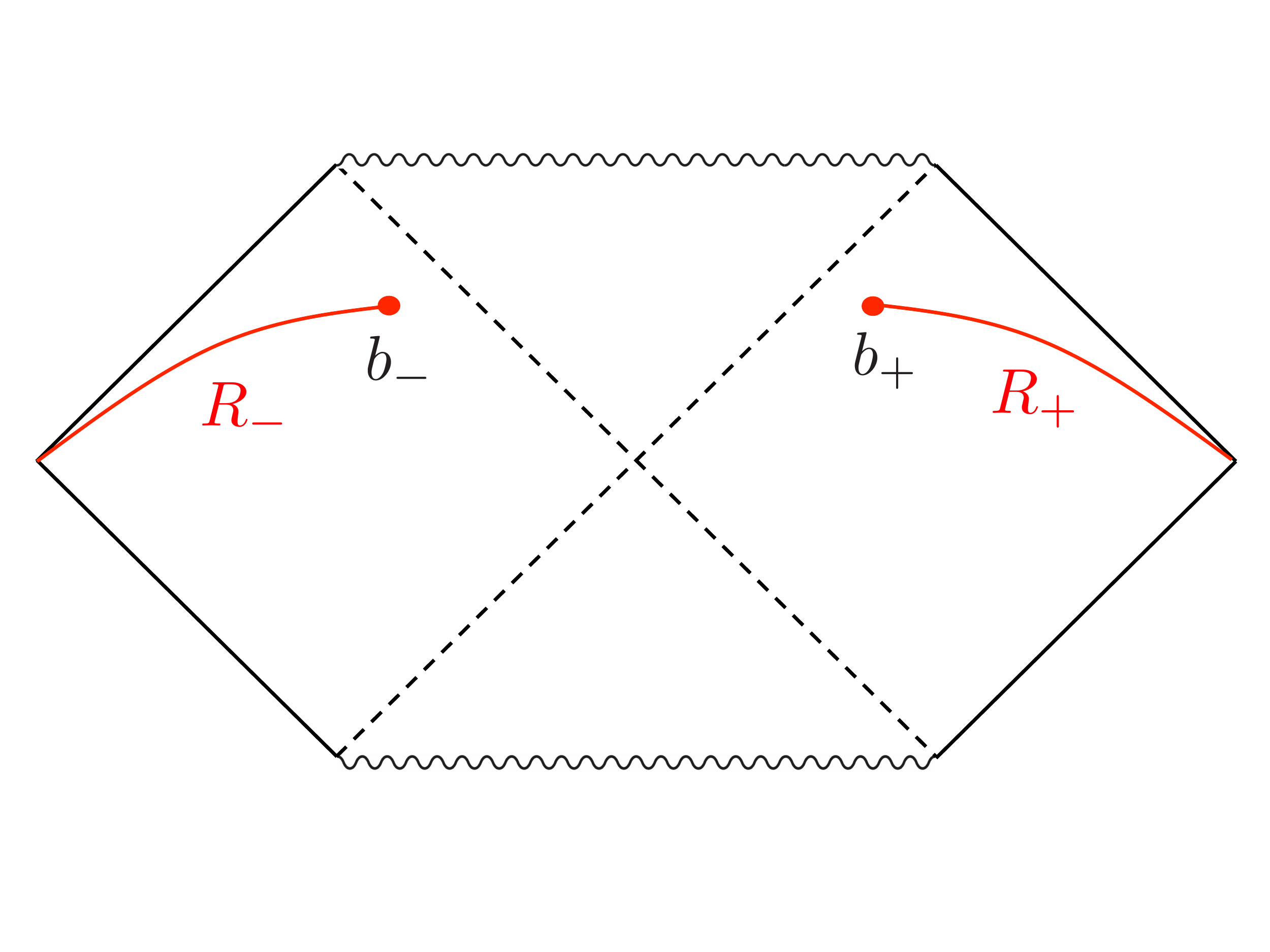}
	\hspace{3mm}
	\includegraphics[width=70mm]{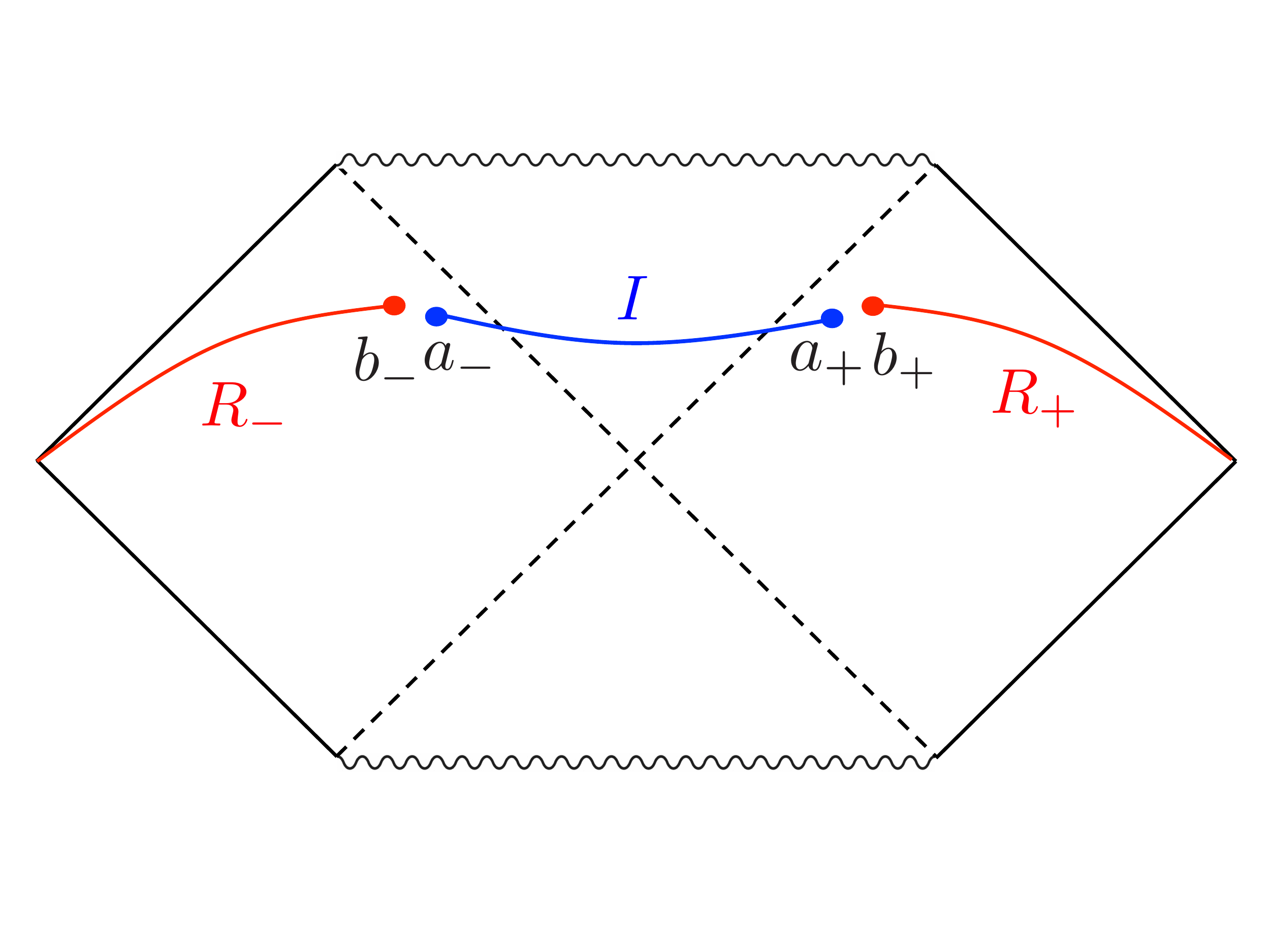}
	\caption{
	Penrose diagram of the static Schwarzschild spacetime 
	without island (left) and that with an island $I$ (right). 
	The region $R$ whose states are identified with the Hawking radiation 
	has two parts $R_+$ and $R_-$, which are located in the right  and the left wedge, respectively. 
	The boundary surfaces of $R_+$ and $R_-$ are indicated as $b_+$ and $b_-$, respectively. 
	The island extends between the right wedge and the left wedge. 
	The boundaries of $I$ are located at $a_+$ and $a_-$. 
	At late times, the distance between the right wedge and the left wedge is very large. 
	}
	\label{fig:BH}
\end{figure}
In fact, 
in the case with an island, a symmetric configuration will give the minimal entropy. 
We may consider only the right wedge,  
since the contributions from the right and left wedges will be equal to each other, so in total it is twice of the 
contribution of only the right wedge. The finite contributions of the matter entanglement entropy from 
only the right wedge is $S(R_+ \cup I)$ minus $S(R_+) + S(I)$, which is essentially the minus of the mutual information $I(R_+;I)$. Therefore, in the case with an island, the finite part of the matter entanglement entropy is given as 
\begin{align}
 \qquad  \qquad  \qquad 
 S^{\rm (finite)}_{\rm matter}(R\cup I)  =  - 2I(R_+,I) 
 \qquad  \qquad  \quad 
 \mbox{(with an island $I$)}
 \label{SwI}
\end{align}
In this paper, for the 
finite matter entropy contribution, $S^{\rm (finite)}_{\rm matter}(R \cup I)$ for eq.~\eqref{GEformula2}, we use eq.~\eqref{SwoI} and eq.~\eqref{SwI}.

The generic expression for the mutual information $I(A;B)$ in curved spacetimes is not known, 
so we need to make some assumptions and take some limits. 
In this paper, we consider only free massless matter fields.
We use the following two limits: the distance between
the boundary surfaces of A and B is (i) large or (ii) small, 
compared to the scale of the size of the boundary surfaces.
\begin{itemize}
\item[(i)]
When the distance is 
much larger than the correlation length of the massive modes 
in the KK tower of the spherical part,
only the s-waves can contribute to $I(A;B)$. 
The mutual information $I(A;B)$ is approximated 
by that of the two-dimensional massless fields,
\begin{equation}
 I(A;B) = - \frac{c}{3} \log d(x,y)
\label{2Dmatter}
\end{equation}
where $c$ is the central charge and $d(x,y)$ is the distance 
between $x$ and $y$ which are the boundaries of $A$ and $B$, respectively. 
\item[(ii)]
When the distance $L$ between the parallelly placed boundary (hyper)surfaces of $A$ and $B$ is
sufficiently small, the mutual information $I(A;B)$ is given by \cite{Casini:2005zv,Casini:2009sr} 
\begin{align}
I(A;B)  = \kappa \,c \, \frac{\mbox{Area}}{L^2}
\label{Ik}
\end{align}
for $c$ free massless matter fields, 
where 
$\kappa$ is a constant.\footnote{
The front numerical factor $\kappa$ in eq.~\eqref{Ik} for a massless field in 4 spacetime dimensions 
is numerically evaluated  \cite{Casini:2009sr} as $\kappa = 0.00554$ (boson) and  $\kappa = 0.00538$ (fermion).
}
Although the formula above is for the flat spacetime,
we expect that it can be used when the length scale of the curvature is large compared to $L$.
\end{itemize}

In this paper, we evaluate the entanglement entropy of 
the Hawking radiation in the asymptotically flat eternal Schwarzschild black hole, and 
investigate the effect of the islands, by using the formulae eqs.~\eqref{GEformula2},  
\eqref{SwoI} and \eqref{SwI} with eqs.~\eqref{2Dmatter} and \eqref{Ik}.
Sec.~\ref{sec:woisland} shows that the entropy without the island grows linearly in time at late times.
In Sec.~\ref{sec:wisland} we include the island and the extremization about the location of it
results in the time-independent behavior of the entropy. We treat two cases: 
Sec.~\ref{sec:close} for the entanglement region $R$ close to the horizon, and Sec.~\ref{sec:far}
for $R$ far away from the horizon. Both cases show that the emergent island extends to the
outer vicinity of the horizon, and the total entropy at late times is almost 
twice the Bekenstein-Hawking entropy formula.
For simplicity the calculations are given for $D=4$ spacetime dimensions, and in Sec.~\ref{sec:higher} we
show that all results are qualitatively the same in generic higher dimensions. 
Finally in Sec.~\ref{sec:summary} we draw the Page curve and discuss the Page time and the scrambling time.
From now on, we write the renormalized Newton constant simply as $G_{\rm N}$.

\section{No island, no entropy bound}
\label{sec:woisland}

In this section we evaluate the entanglement entropy at late times, in the case of the absence of the island.
For the two-dimensional s-wave approximation to make sense, all the points in eq.~\eqref{2Dmatter}
need to be well-separated, compared to the scale of the radius of the sphere. In the absence of the island, we have only two
points which are the boundaries of the entanglement regions at the right $R$ and the left $R$ (see Fig.~\ref{fig:BH} Left). 
So, at late time, we can use the formulas \eqref{SwoI} and 
\eqref{2Dmatter} as
\begin{equation}
 S_{\text{matter}} =  \frac{c}{3} \log d(b_+,b_-)
\end{equation}
where $b_+$ and $b_-$ stand for the boundaries of the entanglement regions in the right and the left wedges of the Schwarzschild geometry.
Here $(t,r) = (t_b,b)$ for $b_+$ and $(t,r) = (-t_b + i {\beta}/{2},b)$ for $b_-$, respectively.\footnote{The imaginary part $i \beta/2$ of time $t$ implies that it is in the left wedge, which means that 
 $U$ and $V$ in the Kruskal coordinates have extra minus sign, as seen in the following.}
Following a conformal map, we find that the matter part of the entanglement entropy in
the Schwarzschild geometry is
\begin{equation}
 S_{\text{matter}} = \frac{c}{6} \log \frac{\left(U(b_-) - U(b_+)\right)\left(V(b_+) - V(b_-)\right)}{W(b_+)W(b_-)} \, .
\end{equation}
Here the Schwarzschild metric in the Kruskal coordinates is given by 
\begin{equation}
 ds^2 = - \frac{dUdV}{W^2} + r^2 d \Omega^2 \ ,
 \label{dUdV} 
\end{equation}
where we have defined the coordinates as
\begin{align}
& r_* = r- r_{\rm h} + r_{\rm h} \log \frac{r-r_{\rm h}}{r_{\rm h}} \ , 
 \\
 & U \equiv - e^{-\frac{t - r_*}{2r_{\rm h}}} = - \sqrt{\frac{r-r_{\rm h}}{r_{\rm h}}} e^{-\frac{t-(r-r_{\rm h})}{2r_{\rm h}}} \, 
 , \quad
 V \equiv e^{\frac{t + r_*}{2r_{\rm h}}} = \sqrt{\frac{r-r_{\rm h}}{r_{\rm h}}} e^{\frac{t+(r-r_{\rm h})}{2r_{\rm h}}} \, .
\end{align}
The conformal factor $W$ of the Schwarzschild black hole geometry is
\begin{align}
W = \sqrt{\frac{r}{4r_{\rm h}}\frac{U V}{r-r_{\rm h}}} = \sqrt{\frac{r}{4r_{\rm h}^3}} \, e^{\frac{r-r_{\rm h}}{2r_{\rm h}}} \ . 
\end{align}
Then, the total entanglement entropy is calculated as 
\begin{equation}
 S = 
\frac{c}{6} \log \left[\frac{16r_{\rm h}^2(b-r_{\rm h})}{b}\cosh^2\frac{t_b}{2 r_{\rm h}}\right] \ . 
\end{equation}

As we mentioned the two-dimensional approximation is valid for late time, $t_b \gg b \, (> r_{\rm h})$, so the above result is 
approximated as
\begin{equation}
 S \simeq
 \frac{c}{6} \frac{t_b}{r_{\rm h}}\,,  
 \label{noisland}
\end{equation}
which grows linearly in time. 

At the late times where 
\begin{align}
\frac{c \, t}{r_{\rm h}} \gg \frac{r_{\rm h}^2}{G_{\rm N}} \,,
\end{align}
this entropy becomes much larger than the black hole entropy. 
This contradicts with the finiteness of the von Neumann entropy for a finite-dimensional black hole system. 
In such a case, an island is expected to emerge. In the next section, we calculate the entanglement entropy with a single island and show that in fact the Page curve is reproduced once we take into account the effects of an island.

\section{Island saves the entropy bound}
\label{sec:wisland}

In this section we calculate the entanglement entropy with a single island. 
The configuration is shown in Fig.~\ref{fig:BH} Right.
To capture the entropy of the full degrees of freedom of the radiation, 
the entanglement region $R$ had better be close to the event horizon.
So first we consider such a case of looking at the black hole closely, $b-r_{\rm h} \ll r_{\rm h}$,  in Sec.~\ref{sec:close}.
Then later in Sec.~\ref{sec:far} we consider the other case when the boundary of the region $R$ is far away from the horizon,
that is, a view from a distance.

\subsection{Close look at the black hole}
\label{sec:close}

Let us consider the case $b-r_{\rm h} \ll r_{\rm h}$, the close look at the black hole from the region $R$.
The boundaries of the island $I$, $a_\pm$, are located at 
$(t,r) = (t_a,a)$ for $a_+$ and $(t,r) = (-t_a + i \beta/2,a)$ for $a_-$. 
It is plausible that $t_a=t_b$ would extremize the entropy, and in this subsection we assume it.
We also assume that at late times, due to the fact that the left wedge and the right wedge are separated
by the volume growing linear in time (see appendix~\ref{sec:app}), we just need to focus on the right-hand side
of the Penrose diagram for the calculation of the entropy (and the final result is twice of it).
Then the total entropy is
\begin{align}
S 
\simeq \frac{2\pi a^2}{G_{\rm N}} - 2 \kappa c \frac{4\pi b^2}{L^2}   \,,
\label{totSis}
\end{align}
where the distance between the end point of the island $I$ and that of the entanglement region $R$ is the 
geodesic distance,
\begin{align}
L = \int_a^b \frac{dr}{\sqrt{1-\frac{r_{\rm h}}{r}}} \, . 
\label{Lis}
\end{align}

We need to extremize the entropy with respect to $a$ which is the location of the boundary of the island.
Physically, when we regard the entropy $S$ as a potential energy for a particle located at $r=a$, this extremization is due to 
the harmonic (gravitational) potential $\frac{2\pi a^2}{G_{\rm N}}$ and the attractive potential $- 2 \kappa c \frac{4\pi b^2}{L^2}$
which pushes the particle closer to $r=b$.
The entropy formula \eqref{Ik} is valid only if $L \ll a$, which is fine because we here consider the case
$b-r_{\rm h} \ll r_{\rm h}$ (and resultantly $a-r_{\rm h} \ll r_{\rm h}$).

In that case, the geodesic distance \eqref{Lis} is
\begin{align}
L \simeq 2\sqrt{r_{\rm h}}
\left(
\sqrt{b-r_{\rm h}}- \sqrt{a-r_{\rm h}}
\right) \, .
\end{align}
To minimize the entropy \eqref{totSis} with respect to $a$, we change the variable to $x \equiv \sqrt{\frac{a-r_{\rm h}}{r_{\rm h}}}$, and consider the equation $\frac{\partial S}{\partial x} = 0$ which is equivalent to  
\begin{align}
x \left(\sqrt{\frac{b-r_{\rm h}}{r_{\rm h}}}-x\right)^3 = \frac{ \kappa c \, G_{\rm N}}{2r_{\rm h}^2} \,, 
\label{eqx}
\end{align}
where approximations $x \ll 1$ and $b \approx r_{\rm h}$ are taken into account. 
This equation has at most two solutions for $x$. The minimization occurs at a smaller $x$ solution, satisfying $x \ll \sqrt{\frac{b-r_{\rm h}}{r_{\rm h}}}$, and with the fact that
the right-hand side of eq.~\eqref{eqx} is very small, we find the location of the island as
\begin{align}
a = r_{\rm h} + \frac{( \kappa c \, G_{\rm N})^2}{4(b-r_{\rm h})^3} \, .
\end{align}
So the boundary of the island is located very close to, and slightly outside of, the black hole horizon.

Substituting this expression to the total entropy, we find 
\begin{align}
S = 
\frac{2 \pi r_{\rm h}^2}{G_{\rm N}}- 2\pi  \kappa \, c \, \frac{r_{\rm h}}{b-r_{\rm h}} \, .
\label{island}
\end{align}
We find a very natural interpretation of this result. First of all, this is constant, as opposed to the late time result without the island,
eq.~\eqref{noisland}. Therefore, the configuration with the island is preferred, and the entropy stops growing at late times.
The first term in   eq.~\eqref{island} 
is exactly (twice of) the Bekenstein-Hawking entropy formula \cite{Hawking:1974sw,Bekenstein:1973ur}. The second term is the effect of the quantum matter.

\subsection{View from a distance}
\label{sec:far}

Next, let us consider the case when the boundary $r=b$ of the entanglement region $R$ is far away from
the horizon, $b \gg r_{\rm h}$. In this case, we {\it assume} that the s-wave approximation is valid,\footnote{
The length scale of the angular sphere at $r=b$ is $b$, while the distance between $a$ ($> r_{\rm h}$) to $b$
is smaller than $b$. So this would cause a problem for adopting the two-dimensional formula \eqref{2Dmatter}. Here,
since the Hawking radiation observed at an asymptotic observer is dominated by the s-wave, we {\it assume} the
use of the two-dimensional formula \eqref{2Dmatter}.
} and use 
the matter entropy formula \eqref{2Dmatter} for calculating the total entropy. 

The entanglement entropy for the conformal matter is given by 
\begin{equation}
 S_{\text{matter}} = \frac{c}{3} \log \frac{d(a_+,a_-) d(b_+,b_-) d(a_+,b_+) d(a_-,b_-)}{d(a_+,b_-) d(a_-,b_+)} \ .
\end{equation}
Using the Kruskal coordinates given in Sec.~\ref{sec:woisland}, the total entanglement entropy is calculated as 
\begin{align}
 S &= \frac{2\pi a^2}{G_{\rm N}} 
 + \frac{c}{6} \log \left[\frac{2^8 r_{\rm h}^4(a-r_{\rm h})(b-r_{\rm h})}{ab}\cosh^2\frac{t_a}{2 r_{\rm h}}\cosh^2\frac{t_b}{2 r_{\rm h}}\right] 
\notag\\&\quad
 + \frac{c}{3} \log \left[\frac{\cosh\left(\frac{r_*(a)-r_*(b)}{2r_{\rm h}}\right)-\cosh\left(\frac{t_a-t_b}{2r_{\rm h}}\right)}
 {\cosh\left(\frac{r_*(a)-r_*(b)}{2r_{\rm h}}\right) + \cosh\left(\frac{t_a+t_b}{2r_{\rm h}}\right)}\right] \ , 
\label{totalE}
\end{align}
where 
\begin{equation}
 \cosh\frac{r_*(a)-r_*(b)}{2r_{\rm h}} 
 = 
 \frac{1}{2} \left[\sqrt{\frac{a-r_{\rm h}}{b-r_{\rm h}}}\,e^{ \frac{a-b}{2r_{\rm h}}} + \sqrt{\frac{b-r_{\rm h}}{a-r_{\rm h}}}\,e^{\frac{b-a}{2r_{\rm h}}} \right] \ . 
 \label{abstar}
\end{equation}
The island is expected to show up near the black hole horizon, so we assume $a \sim r_{\rm h}$ and check if this approximation is correct or not, later. For $a \sim r_{\rm h}$, the second term in the right-hand side of eq.~\eqref{abstar} dominates, so we ignore the first term.

Let us consider the late time behavior. 
We take the late time approximation\footnote{In appendix~\ref{sec:early}, we study the early time behavior and find 
that there is no saddle point for the location of the island, meaning that the island is not generated.}
\begin{equation}
\frac12 \sqrt{\frac{b-r_{\rm h}}{a-r_{\rm h}}} \,e^{ \frac{b-a}{2r_{\rm h}}}
\ll \cosh\frac{t_a + t_b}{2 r_{\rm h}} \, .
\label{latetime}
\end{equation}
We also consider the approximation
\begin{equation}
\cosh\frac{t_a - t_b}{2 r_{\rm h}} \ll
\frac12 \sqrt{\frac{b-r_{\rm h}}{a-r_{\rm h}}} \,e^{ \frac{b-a}{2r_{\rm h}}}
\label{t-t}
\end{equation}
which will be checked to be satisfied later. 
Then the entanglement entropy \eqref{totalE} is approximated as\footnote{
At late times, the distance between the right wedge and the left wedge is very large, so we have
\begin{equation}
 d(b_+,b_-) \simeq d(a_+,a_-) \simeq d(b_\pm,a_\mp) \gg d(b_\pm,a_\pm) \ , 
 \label{LI}
\end{equation}
and the entanglement entropy of the matter is approximated as 
\begin{equation}
 S_{\text{matter}} = \frac{c}{3} \log \left[ d(a_+,b_+) d(a_-,b_-)\right] \ .
\end{equation}
This simplified expression indeed results in the expression same as eq.~\eqref{LS}. 
} 
\begin{align}
 S &= \frac{2\pi a^2}{G_{\rm N}} 
 + \frac{c}{6} \log \left[\frac{2^8 r_{\rm h}^4(a-r_{\rm h})(b-r_{\rm h})}{ab}\cosh^2\frac{t_a}{2 r_{\rm h}}\cosh^2\frac{t_b}{2 r_{\rm h}}\right] 
\notag\\&\quad
 - \frac{c}{3} \log \left[\frac{1}{2}\sqrt{ \frac{a-r_{\rm h}}{b-r_{\rm h}}} \, e^{ \frac{a-b}{2r_{\rm h}}} \cosh\frac{t_a+t_b}{2 r_{\rm h}}\right]
 - \frac{2c}{3} \sqrt{ \frac{a-r_{\rm h}}{b-r_{\rm h}}} \, e^{ \frac{a-b}{2r_{\rm h}}} \cosh\frac{t_a-t_b}{2 r_{\rm h}} \ 
\notag\\
 &= \frac{2\pi a^2}{G_{\rm N}} 
 + \frac{c}{6} \log \left[\frac{16 r_{\rm h}^4 (b-r_{\rm h})^2}{ab} e^{ \frac{b-a}{r_{\rm h}}}\right] 
 - \frac{2c}{3} \sqrt{ \frac{a-r_{\rm h}}{b-r_{\rm h}}} \, e^{ \frac{a-b}{2r_{\rm h}}} \cosh\frac{t_a-t_b}{2 r_{\rm h}} \ . 
 \label{LS}
\end{align}
This allows a local minimum at
\begin{equation}
 a \simeq r_{\rm h} + \frac{(cG_{\rm N})^2}{144\pi^2 r_{\rm h}^2 (b-r_{\rm h})} e^{\frac{r_{\rm h}-b}{r_{\rm h}}} \cosh^2\frac{t_a-t_b}{2 r_{\rm h}} \ , 
 \label{amin}
\end{equation}
and with that value of $a$ the total entropy \eqref{LS} is calculated as 
\begin{align}
 S 
 &= 
 \frac{2\pi r_{\rm h}^2}{G_{\rm N}} 
 + \frac{c}{6} \log \left[\frac{16 r_{\rm h}^3 (b-r_{\rm h})^2}{b} e^{ \frac{b-r_{\rm h}}{r_{\rm h}}}\right] 
 - \frac{c^2 G_{\rm N}}{36\pi r_{\rm h} (b-r_{\rm h})} e^{ \frac{r_{\rm h}-b}{r_{\rm h}}} \cosh^2\frac{t_a-t_b}{2 r_{\rm h}} \ . 
 \label{LSLS}
\end{align}
We vary this expression for $t_a$ and find that $t_a=t_b$ extremizes it. For $t_a=t_b$, the value of $a$ given in eq.~\eqref{amin}
in fact satisfies eq.~\eqref{t-t}. The late time condition \eqref{latetime} is rewritten as 
\begin{align}
r_{\rm h} \log \frac{r_{\rm h}(b-r_{\rm h}) }{c \, G_{\rm N}} \ll t_b \, .
\end{align}
Then in eq.~\eqref{LSLS} we put $t_a=t_b$ and ignore higher order terms in $G_{\rm N}$, to obtain
the final expression of the entanglement entropy
\begin{align}
 S 
 &= 
 \frac{2\pi r_{\rm h}^2}{G_{\rm N}} 
 + \frac{c}{6} 
 \left[
 \log \left(\frac{16 r_{\rm h}^3 (b-r_{\rm h})^2}{b} \right) + \frac{b-r_{\rm h}}{r_{\rm h}}  
 \right]\ . 
 \label{island2}
\end{align}
This does not grow in time.
The interpretation of this result is the same as in the result of the close look, eq.~\eqref{island}.
The first term, which emerged as a result of the island, provides the Bekenstein-Hawking entropy formula
\cite{Hawking:1974sw,Bekenstein:1973ur} for the four-dimensional Schwarzschild black hole.

So summarizing the close look result \eqref{island} and the distant look result \eqref{island2},
we have confirmed that the island shows up at late times and the entropy growth disappears. The boundary of the island is located very close to the event horizon, and in fact the island provides
the renowned Bekenstein-Hawking entropy of the Schwarzschild black hole.


\section{Higher dimensions}
\label{sec:higher}

The arguments presented in this paper can go through for the case of Schwarzschild black holes in higher spacetime dimensions, 
$D\geq 4$. In this section, we provide results in generic $D$ dimensions, and find that the results obtained in Sec.~\ref{sec:woisland}
and Sec.~\ref{sec:wisland} are universal.

The Schwarzschild metric in $D$ dimensions is
\begin{align}
ds^2 = -f(r) dt^2 + \frac{dr^2}{f(r)} + r^2 d\Omega_{D-2}^2\, , \quad f(r) \equiv 1-\frac{r_{\rm h}^{D-3}}{r^{D-3}} \, .
\end{align}
The area of the $(D-2)$-sphere at radius $r$ is $r^{D-2} \Omega_{D-2}$, where $\Omega_{D-2}$ is the volume of the unit $(D-2)$-sphere. 

First, we look at the case with no island.  Similarly to Sec.~\ref{sec:woisland}, the Kruskal coordinates are given\footnote{
The Kruskal coordinates in the right wedge are
\begin{align}
U \equiv -\exp\left[-(D-3)\frac{t-r_*}{2r_{\rm h}}\right] \, , 
\quad
V \equiv \exp\left[(D-3)\frac{t+r_*}{2r_{\rm h}}\right] \, , 
\end{align}
with $r_* \equiv \int^r dr /f(r)$, giving the metric of the form \eqref{dUdV} with
\begin{align}
W \equiv \frac{(D-3)}{2r_{\rm h}\sqrt{f(r)}} \exp\left[ (D-3)\frac{r_*}{2r_{\rm h}}\right] \, .
\end{align}
}
just by generalizing 
the factor $f(r)$. We arrive at the expression
for the total entropy at late times,
\begin{align}
S \simeq 
\frac{c}{6}(D-3)\frac{t_b}{r_{\rm h}} \, .
\label{noislandh}
\end{align}
For $D=4$ this reproduces eq.~\eqref{noisland}. There is no physical difference; the entropy grows linearly in time.

Next, we consider the entanglement entropy with the island. For the close look at the black hole as in 
Sec.~\ref{sec:close},
instead of the four-dimensional formula \eqref{Ik}, we use the $D$-dimensional formula
\begin{align}
I(A;B) = \kappa_D \, c \, \frac{\mbox{Area}}{L^{D-2}} \, .
\end{align}
Here the constant $\kappa_D$ also depends on $D$. 
The total entropy with the island contribution is
\begin{align}
S \simeq \frac{\Omega_{D-2}}{2G_{\rm N}} a^{D-2} -2 \kappa_D \, c\frac{\Omega_{D-2} \,b^{D-2}}{L^{D-2}} 
\end{align}
with the distance in the short distance approximation $b,a\simeq r_{\rm h}$, 
\begin{align}
L \simeq \frac{2}{\sqrt{D-3}}\sqrt{r_{\rm h}}\left(\sqrt{b-r_{\rm h}}-\sqrt{a-r_{\rm h}}\right)\, .
\end{align}
We minimize $S$ by varying the location $a$ of the boundary of the island, and find
\begin{align}
a = r_{\rm h} + \frac{(\kappa_D\,  c \, G_{\rm N})^2}{r_{\rm h}^{2D-5}}  2^{6-2D} (D-3)^{D-2} 
\left(\frac{r_{\rm h}}{b-r_{\rm h}}\right)^{D-1} \, .
\end{align}
So the island is located very close to the black hole horizon. The total entanglement entropy is found as
\begin{align}
S = \frac{\Omega_{D-2}}{2G_{\rm N}} r_{\rm h}^{D-2}
-\Omega_{D-2} \kappa_D \,  c \, 2^{3-D} (D-3)^{(D-2)/2} \left(\frac{r_{\rm h}}{b-r_{\rm h}}\right)^{(D-2)/2} \, .
\label{higherSclose}
\end{align}
This reproduces eq.~\eqref{island} for $D=4$. Again, the contribution of the Bekenstein-Hawking entropy emerges
as the island contribution, and there exists a small contribution from the matter field (the second term).

We can also work out the higher dimensional case for the view at a distance given in Sec.~\ref{sec:far}. 
The total entanglement entropy with the island at late times is
\begin{align}
S \, \simeq \,  &
\frac{\Omega_{D-2}}{2G_{\rm N}} a^{D-2}
+\frac{c}{6}
\left[
\log\left[\frac{2^4 r_{\rm h}^4 f(b)f(a)}{(D-3)^4}\right] + 2(D-3)\frac{r_*(b)-r_*(a)}{2r_{\rm h}}
\right]
\nonumber
\\
& -\frac{2c}{3}\exp\left[ (D-3)\frac{r_*(a)-r_*(b)}{2r_{\rm h}}\right] \cosh \left[(D-3)\frac{t_a-t_b}{2r_{\rm h}}\right] \, .
\end{align}
The location of the boundary of the island is again found to be very close to the horizon,
\begin{align}
a = r_{\rm h}
+ \frac{(c\, G_{\rm N})^2}{r_{\rm h}^{2D-5}}\left(\frac{2}{3(D\!-\!2)\Omega_{D\!-\!2}}\right)^2 \!
\exp\!\left[
(3\!-\! D)\left(\frac{b}{r_{\rm h}}\!-\!1\! +\! g\!\left(\!\frac{b}{r_{\rm h}}\!\right)\right)
\right]
 \cosh^2 \!\left[(D\!-\!3)\frac{t_a-t_b}{2r_{\rm h}}\right] \, ,
 \label{ahigh}
\end{align}
where we have defined\footnote{The function $g(x)$ reduces to $\log(x-1)$ at $D=4$. The complicated form of $g(x)$ is just because of 
the Kruskal coordinates in higher dimensions.} 
\begin{align}
g(x)\!\equiv\! \frac{-x^{4-D}}{D\!-\!4} {}_2F_1 \! \left(\!1,\frac{D\!-\!4}{D\!-\!3},\frac{2D\!-\!7}{D\!-\!3};x^{3-D}\!\right)
-\frac{1}{D\!-\!3}\left(\!\gamma\! +\! \log(D\!-\!3) \!+\! \frac{\Gamma'((D\!-\!4)/(D\!-\!3))}{\Gamma((D\!-\!4)/(D\!-\!3))}\right) \, .
\nonumber
\end{align}
Substituting eq.~\eqref{ahigh} to the total entropy, we find again $t_a=t_b$ extremizes it, and the final expression for the entanglement
entropy is 
\begin{align}
S \simeq
\frac{\Omega_{D-2}}{2G_{\rm N}} r_{\rm h}^{D-2}
+\frac{c}{6}
\left[
\log\!\left[\frac{2^4 r_{\rm h}^4 f(b)}{(D\!-\!3)^3}\right] + (D\!-\!3)\left(\frac{b}{r_{\rm h}}\!-\!1\! +\! g\!\left(\!\frac{b}{r_{\rm h}}\!\right)\right)
\right] \, .
\label{higherSfar}
\end{align}
The first term of this late-time expression of the entanglement entropy is the Bekenstein-Hawking entropy which emerged from
the island.
This eq.~\eqref{higherSfar} shares the same structure as eq.~\eqref{higherSclose}.
 

\section{Page time and scrambling time}
\label{sec:summary}

In this paper, we have calculated the entanglement entropy of the Hawking radiation of the asymptotically flat eternal Schwarzschild black hole
in $D$ ($\geq 4$) spacetime dimensions, 
for the configuration without islands and that with an island. 
We can summarize the findings, eqs.~\eqref{noisland}, \eqref{noislandh}, 
\eqref{island}, \eqref{island2}, \eqref{higherSclose}, and \eqref{higherSfar}, as follows. 
The entanglement entropy of a given region $R$ outside of the horizon
linearly grows with time $t$
for the configuration without islands; 
\begin{equation}
 S = \frac{c}{6}(D-3) \frac{t}{r_{\rm h}}  \,, 
 \label{BHSt} 
\end{equation}
where $r_{\rm h}$ is the Schwarzschild radius and $c$ is the number of massless matter fields.
For the case with an island at late times, 
the saddle point analysis for 
the boundary of the island $a$ shows that it emerges at the outer vicinity of the horizon, 
\begin{align}
a = r_{\rm h} + {\cal O}\left(\frac{(c \, G_{\rm N})^2}{r_{\rm h}^{2D-5}}\right) \, .
\label{arh}
\end{align}
The resultant entanglement entropy for the region $R$ is\footnote{
In this paper we have assumed the dimensionless quantity $c \, G_{\rm N}/r_{\rm h}^{D-2} \ll 1$. 
This applies to our universe, since the number of massless fields $c$ is not very large, 
while the Newton constant $G_{\rm N}$, or equivalently, the Planck length is 
much smaller than the typical scale of realistic black holes. 
Therefore, the higher order terms in eq.~\eqref{arh} and similar expressions are negligible. 
} 
\begin{align}
S = 
2 \, S_{\rm BH} + {\cal O}(c) \, ,
\label{BHres}
\end{align}
where $S_{\rm BH}$ is the Bekenstein-Hawking entropy of the Schwarzschild black hole, 
$S_{\rm BH} \equiv \mbox{Area} (r\!=\!r_{\rm h})/4G_{\rm N}$, 
which is time-independent at late times. $ {\cal O}(c)$ effects arise from the quantum effects by the matters. 

\begin{figure}[t]
\centering
	\includegraphics[width=80mm]{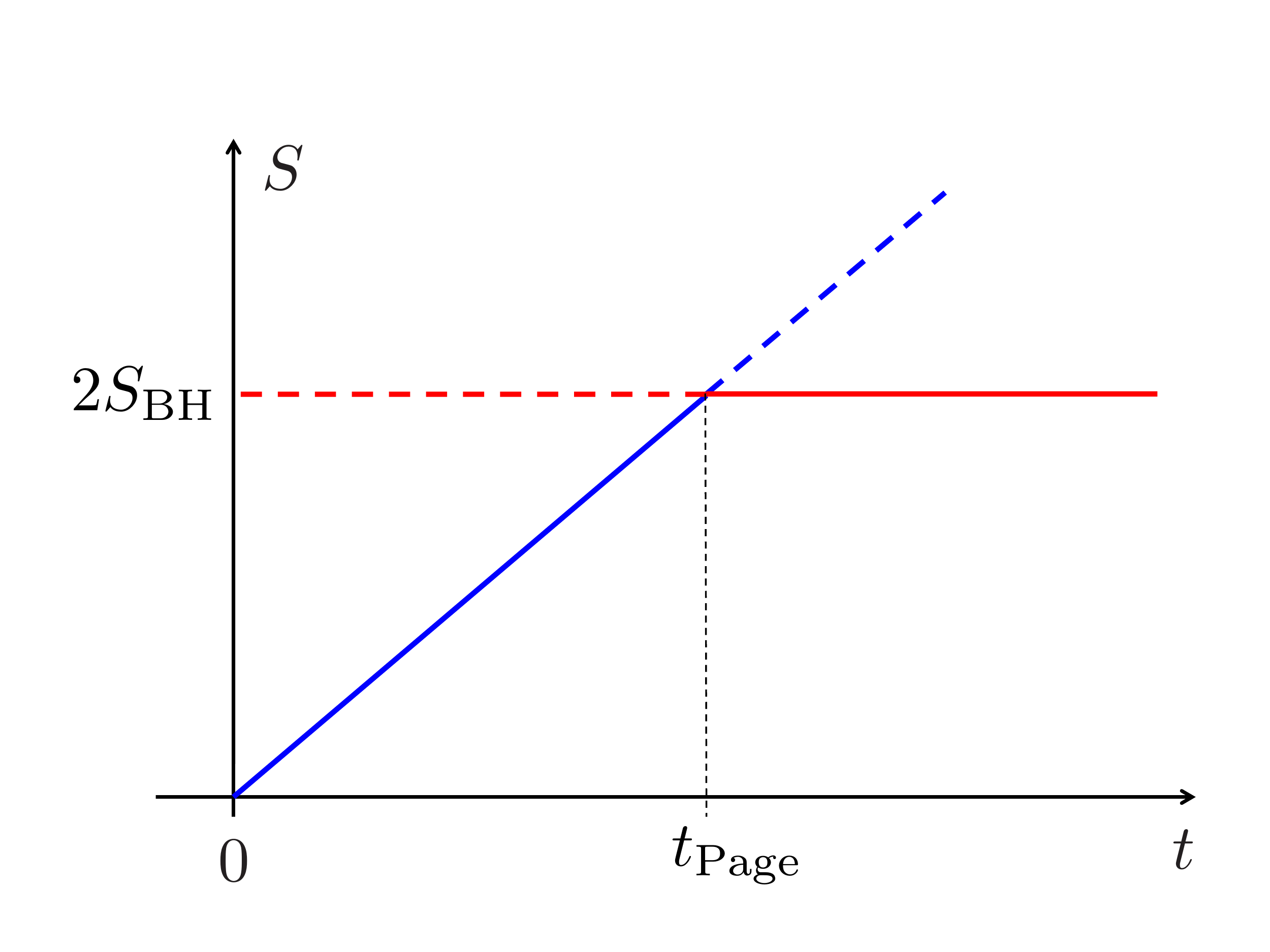}
	\caption{
	The Page curve for the eternal Schwarzschild black hole. In this plot we ignore terms of higher order in $c \, G_{\rm N}/r_{\rm h}^{D-2}$, which are small compared to $t_{\rm Page}$ or $S_{\rm BH}$.
	}
	\label{fig:Page}
\end{figure}

Generally, the dominant contribution to the entanglement entropy comes from 
the configuration with a minimum entropy. 
Thus, for the eternal Schwarzschild black holes, at early times the entanglement entropy is given by that of the configuration without the island, 
then at late times it is by the one with the island. 
So the dominant configurations switch at the time identified with Page time $t_{\rm Page}$, 
at which the time evolution of the entanglement entropy drastically changes: 
the linear growth is replaced by a time-independent constant, see Fig.~\ref{fig:Page}.
Equating the asymptotic constant value of the entropy \eqref{BHres} 
with the entropy without the island \eqref{noislandh}, we find the Page time for the eternal Schwarzschild black hole,
\begin{align}
t_{\rm Page} = \frac{3\Omega_{D-2}}{(D-3)} \frac{r_{\rm h}^{D-1}}{c \, G_{\rm N}} + \mathcal O\left(r_{\rm h}\right) \,.
\end{align}
Although the higher order corrections would depend on $b$, 
which is the boundary location of the entanglement region $R$, 
the leading term is universal. 
Using the Hawking temperature $T_{\rm H} = \frac{(D-3)}{4 \pi r_{\rm h}}$, the universal term is written as
\begin{align}
t_{\rm Page} = \frac{3}{\pi} \frac{S_{\rm BH}}{c \, T_{\rm H}}  \, .
\label{Pagetime}
\end{align}
After the Page time, the entropy is given by $2 S_{\rm BH}$ 
for the Hawking radiation in the both sides of the Penrose diagram. 
Thus the entanglement entropy for the Hawking radiation observed only in a single side approximately agrees with $S_{\rm BH}$, as expected.


Let us compare the Page time with a semiclassical estimate of the lifetime of the black hole \cite{Page:1976df}.
In four dimensions, the radiation power reduces the mass $M$ of the black hole as 
\begin{align}
\frac{dM}{dt} = -\frac{c \alpha}{G_{\rm N}^2 M^2}
\end{align}
where $\alpha$ is a constant dependent on the spin of the radiating particle. 
Solving this gives a time-dependent Schwarzschild radius as 
\begin{align}
r_{\rm h}(t) = r_{\rm h}(t=0) \left[1-24 c \alpha \frac{G_{\rm N} \, t}{(r_{\rm h}(t=0))^3}\right]^{1/3} \, , 
\label{rt}
\end{align}
so the semiclassical estimate of the black hole lifetime is 
\begin{align}
t_{\rm evaporate} = \frac{(r_{\rm h}(t=0))^3}{24 c\,  \alpha G_{\rm N}} = \frac{1}{96 \pi^2 \alpha} 
\frac{S_{\rm BH}(t=0)}{c \, T_{\rm H}(t=0)} \, .
\label{evapo}
\end{align} 
On the other hand, the Page time \eqref{Pagetime} can be modified once we include this semiclassical reduction of the 
black hole mass. 
Since the contribution of the change of the geometry to the matter entropy is at higher order, we can just 
substitute the time-dependent Schwarzschild radius \eqref{rt} to our entropy formulas eqs.~\eqref{BHSt} and \eqref{BHres}, to evaluate
the Page time with the effect of the black hole evaporation. 
At the early stage, the time evolution of the entropy slightly deviates from the linear growth. And at late times, the
entropy is not constant but decreases in time. The intersection of the two curves gives the Page time. 
We find, in four dimensions, the Page time is obtained by solving 
\begin{align}
\frac{c}{6} \frac{t}{r_{\rm h}(t)}   = 2 \frac{4 \pi \left( r_{\rm h}(t) \right)^2}{4G_{\rm N}}  \,,
\end{align}
which yields
\begin{align}
t_{\rm Page} = \frac{3}{\pi} \frac{S_{\rm BH}(t=0)}{c \, T_{\rm H}(t=0)}  \frac{1}{1 + 2^5 3^2 \pi \alpha} \, .
\label{Pagemo}
\end{align}
Looking at the original eq.~\eqref{Pagetime}, we find that 
the last factor in the expression above is due to the backreaction of the black hole evaporation.
Comparing this Page time \eqref{Pagemo} with the black hole lifetime \eqref{evapo}, we see that both are proportional
to $S_{\rm BH}(t=0) / (c \, T_{\rm H}(t=0))$, so they are at the same order.\footnote{
The numerical ratio of the two, in the s-wave approximation which provides 
$\alpha = \frac{1}{384\pi}$, is found as $t_{\rm Page}/t_{\rm evaporate} = 0.43$. 
(Note that the value of $\alpha$ is a bit larger than that in ref.~\cite{Page:1976df} 
as the effect of the graybody factor is not taken into account in this paper.)
One can compare this with the estimate in ref.~\cite{Page:2013dx} which gives the ratio $\sim 1/2$.}

With the concrete location of the emergent island, we can also discuss the time scale for scrambling.
According to the island prescription, 
the density matrix of the Hawking radiation in $R$ is 
effectively given by that of $R$ and $I$. 
This implies that the information thrown into the island $I$ 
would be able to be collected from the Hawking radiation. 
If we send a message from the point $r=b$ toward the island at time $t=t_0$, 
it reaches the island $r=a$, at time
\begin{equation}
 t_a = t_0 + b - a + r_{\rm h} \log \frac{b - r_{\rm h}}{a - r_{\rm h}} \ , 
\end{equation}
at the earliest.\footnote{For simplicity the formula is written for the $D=4$ case.} Supposing that the message would be 
reconstructable from the Hawking radiation once they are in the island, $I$ at $t_b = t_a$, 
we can identify the scrambling time $t_{\rm scr} = t_a - t_0$, 
since the information is no longer contained in the black hole 
but in the Hawking radiation $R \cup I$. 
Using eq.~\eqref{arh}, this yields the scrambling time estimation as 
\begin{align}
t_{\rm scr}  \, \simeq  \, 2 r_{\rm h}\log \frac{ r_{\rm h}^2 }{ G_{\rm N}} \, \simeq  \, \frac{1}{2\pi T_{\rm H}} \log S_{\rm BH} \, .
\end{align}
Since the leading contribution comes from $\log G_{\rm N}$, 
the most dominant part of the scrambling time $t_{\rm scr}$ is universal 
and expressed in terms of the Hawking temperature $T_{\rm H}$ and 
the Bekenstein-Hawking entropy $S_{\rm BH}$.  
This expression is valid in general dimensions, $D\geq 4$. 
The scrambling time obtained is proportional to $1/T_{\rm H} \log S_{\rm BH}$, as predicted first in ref.~\cite{Sekino:2008he}.

Several comments are in order.
In this paper, we have studied only the configuration without islands and that with an island. 
Configurations with more islands also might contribute to the entanglement entropy. 
As the configuration with a single island agrees with the entropy of the black hole, 
those with more islands would not have dominant contributions at late times. 
They would contribute around the Page time, so that the sharp change of the 
time evolution of the entanglement entropy may be smoothed away. 

One remaining problem, which is important in the viewpoint of information, is how 
the information in the island is transported to the Hawking radiation. 
We have found that the expected Page curve is reproduced by the effect of the island, 
and the entanglement entropy of the Hawking radiation agrees with that of the black hole.
However these do not tell how the information is restored concretely. Further study 
of islands will reveal the mystery of the black hole information paradox.

\acknowledgments

K.~H.~would like to thank Sotaro Sugishita for discussions on ref.~\cite{Carmi:2017jqz}. 
N.~I.~would like to thank Takanori Anegawa for various helpful discussions on related project \cite{Anegawa:2020ezn}. 
This work is supported in part by JSPS KAKENHI Grant Number JP17H06462 (K.H.), JP18K03619 (N.I.).

\appendix

\section{Early time growth of the entropy}
\label{sec:early}

In this appendix we study the early time growth of the entanglement entropy.
At the very early stage of the time evolution, the geodesic distance between the two points is short compared to the 
scale of the area, so we can use the short distance formula \eqref{Ik}. 

In the formula \eqref{Ik} the separation $L$ between the two regions $R$ is approximated by the geodesic distance 
in two-dimensional part of the Schwarzschild spacetime. At the short distance expansion it should coincide\footnote{We
check indeed it coincides, see appendix \ref{sec:app}.} with $d$ in
Sec.~\ref{sec:woisland}, 
\begin{align}
L \simeq 4 r_{\rm h} \sqrt{\frac{b-r_{\rm h}}{r_{\rm h}}} \cosh \frac{t}{2r_{\rm h}}\, .
\label{Lshort}
\end{align}
For this $L$ to be smaller than the scale of the area $\sim r_{\rm h}^2$, we need $t \ll r_{\rm h} \log \frac{r_{\rm h}}{b-r_{\rm h}}$
and thus $b-r_{\rm h} \ll r_{\rm h}$.

Substituting eq.~\eqref{Lshort} to eq.~\eqref{Ik}, we find the total entropy,
\begin{align}
S = 
-  \frac{\pi^2 \kappa c}{4 \left(\frac{b}{r_{\rm h}}-1\right) \cosh^2 \frac{t}{2 r_{\rm h}}} \, . 
\end{align}
This grows as $\sim t^2$ at early time $0\leq t \ll r_{\rm h} \log \frac{r_{\rm h}}{b-r_{\rm h}}$.

So, together with the result given in Sec.~\ref{sec:woisland}, 
we find that the total entanglement entropy grows in time, when we do not include the 
contribution from the island. At early times it grows as $t^2$, and at late times it grows linearly in time.

Now, let us consider if we need to include the contribution from the island, at early times. 
We look at the early time behavior of eq.~\eqref{totalE}.\footnote{
For the two-dimensional approximation to work, the geodesic distance between $a_-$ and $a_+$ needs to be large compared to $r_{\rm h}$.
Looking at the formula \eqref{Lshort}, this means $\cosh t_a \gg \sqrt{\frac{a-r_{\rm h}}{r_{\rm h}}}$. When $a$ is close
enough to $r_{\rm h}$, this is satisfied for any $t_a$.
}
Suppose 
\begin{equation}
\frac12 \sqrt{\frac{b-r_{\rm h}}{a-r_{\rm h}}} \,e^{ \frac{b-a}{2r_{\rm h}}}
\gg \cosh\frac{t_a + t_b}{2 r_{\rm h}} \, , \, \cosh\frac{t_a - t_b}{2 r_{\rm h}} \ 
 \label{et}
\end{equation}
which is valid at early times, $t_a, t_b \ll r_{\rm h}$.
The entanglement entropy \eqref{totalE} is approximated as 
\begin{align}
 S &= \frac{2\pi a^2}{G_{\rm N}} 
 + \frac{c}{6} \log \left[\frac{2^8 r_{\rm h}^4(a-r_{\rm h})(b-r_{\rm h})}{ab}\cosh^2\frac{t_a}{2 r_{\rm h}}\cosh^2\frac{t_b}{2 r_{\rm h}}\right] 
\notag\\&\quad
 - \frac{4c}{3} \sqrt{ \frac{a-r_{\rm h}}{b-r_{\rm h}}} \, e^{ \frac{a-b}{2r_{\rm h}}} 
 \cosh\frac{t_a}{2 r_{\rm h}}\cosh\frac{t_b}{2 r_{\rm h}} \ .
 \label{Sshort}
\end{align}
The location of the boundary of the island, $r=a$, is determined by minimization of eq.~\eqref{Sshort}. However, we find no
saddle point, as seen in the following way. Writing $a$-dependent terms in eq.~\eqref{Sshort} by 
$x \equiv \sqrt{(a-r_{\rm h})/r_{\rm h}}$,
eq.~\eqref{Sshort} is roughly written as 
\begin{align}
S \sim \frac{r_{\rm h}^2}{G_{\rm N}} x^2 + c \log x - c x \, .
\end{align}
So, the saddle point equation has the structure
\begin{align}
0 = \frac{\partial S}{\partial x} = \frac{r_{\rm h}^2}{G_{\rm N}} x + \frac{c}{x} - c  \, .
\end{align}
For $G_{\rm N} c/r_{\rm h}^2 \ll 1$, this does not allow a solution for $x$. 
Since there is no minimum for the entropy as we vary $a$, we conclude that at early times the island is not generated.


\section{Geodesic distance and extremal volume}
\label{sec:app}

In appendix~\ref{sec:early}, we have argued that 
the length scale $d$ used in Sec.~\ref{sec:woisland} is equal to 
the separation $L$ between the two regions $R$, given by the geodesic distance 
in the two-dimensional part of the Schwarzschild spacetime, 
\begin{equation}
 L = \int dt \sqrt{-\left(1-\frac{r_{\rm h}}{r}\right) + \left(1-\frac{r_{\rm h}}{r}\right)^{-1} \dot r^2} \ . 
 \label{gd}
\end{equation}
Here we show that it is indeed the case.

By integrating the equation of motion of eq.~\eqref{gd}, or equivalently, 
since the Hamiltonian must be constant, 
we obtain a conservation law 
\begin{equation}
 \lambda = \frac{(1-\frac{r_{\rm h}}{r})}{\sqrt{-\left(1-\frac{r_{\rm h}}{r}\right) + \left(1-\frac{r_{\rm h}}{r}\right)^{-1} \dot r^2}} \, , 
\label{eom}
\end{equation}
where $\lambda$ is a constant. By solving eq.~\eqref{eom}, we find $dt/dr$ as
\begin{equation}
 \frac{dt	}{dr} = \frac{1}{\left(1-\frac{r_{\rm h}}{r}\right)\sqrt{1+\lambda^{-2}\left(1-\frac{r_{\rm h}}{r}\right)}} \ . 
\end{equation}
We refer the ``reversing'' point $\dot r = 0$ as $(0,r_0)$, then we have 
\begin{equation}
 t  = \int_{r_0}^{b} \frac{dr}{\left(1-\frac{r_{\rm h}}{r}\right)}
 \sqrt{\frac{\left(1-\frac{r_0}{r_{\rm h}}\right)}{\left(1-\frac{r_0}{r}\right)}} \ , 
 \label{tr}
\end{equation}
and the integration constant is determined as $\lambda^2 = \frac{r_{\rm h}}{r_0} - 1$.
Now, early time corresponds to $r_0 \simeq r_{\rm h}$, so we define $r_0 = r_{\rm h}(1+\alpha)$ and $b=r_{\rm h}(1+\beta)$,
and consider the parameter regions $\alpha\ll 1 $ and $\beta \ll 1$. The latter is natural as we need a shorter distance.
In this approximation, we find that eq.~\eqref{tr} is solved for a relation between $\alpha$ and $t$ as
\begin{align}
\beta-\alpha = \beta \cosh^2 \frac{t}{2 r_{\rm h}} \, . 
\label{t}
\end{align}
The geodesic distance \eqref{gd} to which the expression for $\dot{r}$ is substituted, is 
\begin{align}
L = 2 \int_{r_0}^b dr \sqrt{\frac{r_0}{r_{\rm h}}} \frac{1}{\sqrt{1-\frac{r_0}{r}} }\, .
\label{lr0}
\end{align}
Since $0<r_0 < r_{\rm h} < b$, the integrands above are divergent at $r=r_{\rm h}$, but the divergence can be regularized
by cutting out the region $r_{\rm h} - \epsilon < r < r_{\rm h} + \epsilon$ with $\epsilon \ll r_{\rm h}$. 
In the approximation $\beta \ll 1$, we find
\begin{align}
L \, \simeq \, 4 r_{\rm h} \sqrt{\beta-\alpha} = 4 r_{\rm h} \sqrt{\beta} \cosh \frac{t}{2r_{\rm h}}\, .
\label{Lshort2}
\end{align}
In the second equality we substituted eq.~\eqref{t}.
This indeed coincide with eq.~\eqref{Lshort}.

Another issue which we would like to settle here is the assumption we have used in Sec.~\ref{sec:close}: the growth of 
the volume $V(t)$ of the extremal surface between the boundary at $b_+$ and that at $b_-$. The calculation of $V(t)$ is
quite similar to that of the holographic complexity \cite{Carmi:2017jqz}, 
except that we are now working in the Schwarzschild spacetime.
Following the method developed in ref.~\cite{Carmi:2017jqz}, after a straightforward calculations similar to the above,
we find
\begin{align}
V = \int_{r_0}^{b}dr \frac{r^4 }{\sqrt{r^4 f(r) - r_0^4 f(r_0)}} \, , \quad
t = \int_{r_0}^{b}dr \frac{-\sqrt{-r_0^4f(r_0)} }{f(r) \sqrt{r^4 f(r) - r_0^4 f(r_0)}} \, .
\end{align}
with $f(r)\equiv 1-r_{\rm h}/r$. If we drop $r^4$ and $r_0^4$ from the expression above, they reduce to eq.~\eqref{lr0} and eq.~\eqref{tr}.
These equations give $V(r_0)$ and $t(r_0)$ where $r_0$ is the value of $r$ at the ``reversing'' point. So, eliminating $r_0$,
we obtain $V(t)$. Numerically, we can easily find that $V(t)$ grows linearly in time at late times, as in the case of
the holographic complexity \cite{Carmi:2017jqz}.


\end{document}